\documentclass[apj]{emulateapj}

\usepackage{color}

\begin{document}


\title{Observational Evidence of a Flux Rope within a Sunspot Umbra}


%

\author{Salvo L. Guglielmino\altaffilmark{1}, Paolo Romano\altaffilmark{2} and Francesca Zuccarello\altaffilmark{1}}
\email{salvo.guglielmino@oact.inaf.it}

\altaffiltext{1}{Dipartimento di Fisica e Astronomia - Sezione Astrofisica, Universit\`{a} di Catania,
			 Via S. Sofia 78, I-95125 Catania, Italy}
\altaffiltext{2}{INAF - Osservatorio Astrofisico di Catania,
	Via S. Sofia 78, I-95125 Catania, Italy}

\begin{abstract}

We observed an elongated filamentary bright structure inside the umbra of the big sunspot in active region NOAA 12529, which differs from the light bridges usually observed in sunspots for its morphology, magnetic configuration, and velocity field. We used observations taken with the \textit{Solar Dynamic Observatory} satellite to characterize this feature. Its lifetime is 5 days, during which it reaches a maximum length of about 30\arcsec{}. In the maps of the vertical component of the photospheric magnetic field, a portion of the feature has a polarity opposite to that of the hosting sunspot. At the same time, in the entire feature the horizontal component of the magnetic field is about 2000~G, substantially stronger than in the surrounding penumbral filaments. Doppler velocity maps reveal the presence of both upward and downward plasma motions along the structure at the photospheric level. Moreover, looking at the chromospheric level, we noted that it is located in a region corresponding to the edge of a small filament which seems rooted in the sunspot umbra. Therefore, we interpreted the bright structure as the photospheric counterpart of a flux rope touching the sunspot and giving rise to penumbral-like filaments in the umbra.

\end{abstract}



\section{Introduction}

High resolution observations acquired by the new generation of ground-based and space-borne telescopes are allowing us to study more and more details of the photospheric fine structure that characterizes sunspots \citep{Sol03, Sch09}. In particular, the investigation of the bright structures observed inside the dark umbrae, like umbral dots or light bridges (LBs), has become essential to understand the physical mechanisms that are responsible for the heat transport from the convection zone into the photosphere and for the diffusion of the magnetic field. 

LBs are structures that rapidly intrude from the leading edge of penumbral filaments into the umbra. They are often linked to fragmentation of sunspots and to the final phases of the sunspot evolution \citep[e.g.,][and references therein]{Fal16,Fel16}, although not all spots exhibit this phenomenon. \citet{Sob97} classified LBs into strong or faint, depending on whether they separate the umbra into distinct umbral cores. Moreover, he distinguished between granular or filamentary LBs, depending on their internal structure. In particular, filamentary LBs have a central dark lane, i.e., a main dark axis along the LB, and branches of dark and bright hairs connected to the central lane \citep[e.g.,][]{Lit04}. Another classification was proposed by \citet{Tho04}, who distinguished between segmented and unsegmented LBs. The former, which are by far the most common, are characterized by bright granules separated by dark lanes perpendicular to the LB axis \citep{Ber03}, although the granular pattern visible in these structures differs in size, lifetime, and brightness from the quiet Sun granulation \citep[see, e.g.,][]{Lag14,Fal17}. On the contrary, unsegmented LBs resemble the elongated filaments forming the penumbra of sunspots, without any evidence of granular cells. In recent observations performed at the GREGOR telescope, \citet{Sch16} identified a new class of LBs, called thick LBs, with small transverse \textsc{Y}-shaped dark lanes similar to dark-cored penumbral filaments. 

Doppler velocity measurements show the presence of upward plasma motions above LBs at the photopheric level, specially along the dark lane \citep{Rim97, Gio08, Rou10}. It seems that the plasma convection takes the lead in these regions, which are characterized by a magnetic field weaker and more horizontal than in the surrounding umbra \citep{Lit91, Lek97, Jur06}. However, it is still controversial whether LBs have a magneto-convective origin or are due to the field-free convection that penetrates into the strong umbral magnetic field from below the photosphere and forms a cusplike magnetic field near the visible surface \citep{Tho04, BI11}. Remarkable long-lasting plasma ejections or surge activities are observed in the chromosphere along some LBs \citep[e.g.][]{Bha07, Shi09, Lou14, Tor15}. Magnetic reconnection occurring in the low chromosphere and the upper photosphere is thought to originate this small-scale eruptive activity above LBs \citep{Son17}.

Recently, \citet{Kle13} analyzed unusual filamentary structures observed within the umbra of the very flare-productive active region (AR) NOAA 11302 preceding sunspot. These structures, called umbral filaments (UFs), do not resemble typical LBs in morphology or in evolution and are formed by curled filaments that reach from the penumbra well into the umbra. Furthermore, UFs show counter Evershed flow along them at the photospheric level and energy dissipation phenomena in the higher atmospheric layers.

In this Letter, we describe the observation of an elongated bright structure inside the umbra of the big preceding sunspot of AR NOAA 12529. Curiously, its appearance transformed the original spot into a huge, heart-shaped sunspot, so that it became celebrated in popular media. We investigated this feature, which at first glance resembles a filamentary LB. We find that it is characterized by mixed directions of the plasma motions and a strong horizontal magnetic field, with a portion of the structure having opposite polarity to that of the hosting sunspot. These and other signatures allow us to interpret this structure as a manifestation of a flux rope that is located in higher layers of the solar atmosphere above the umbra. We describe the data set and its analysis in Sect.~2. The observations are presented in Sect.~3 and our conclusions are reported in Section~4.

\begin{figure}[t]
	\centering
	\includegraphics[trim=10 125 120 480, clip, scale=0.6]{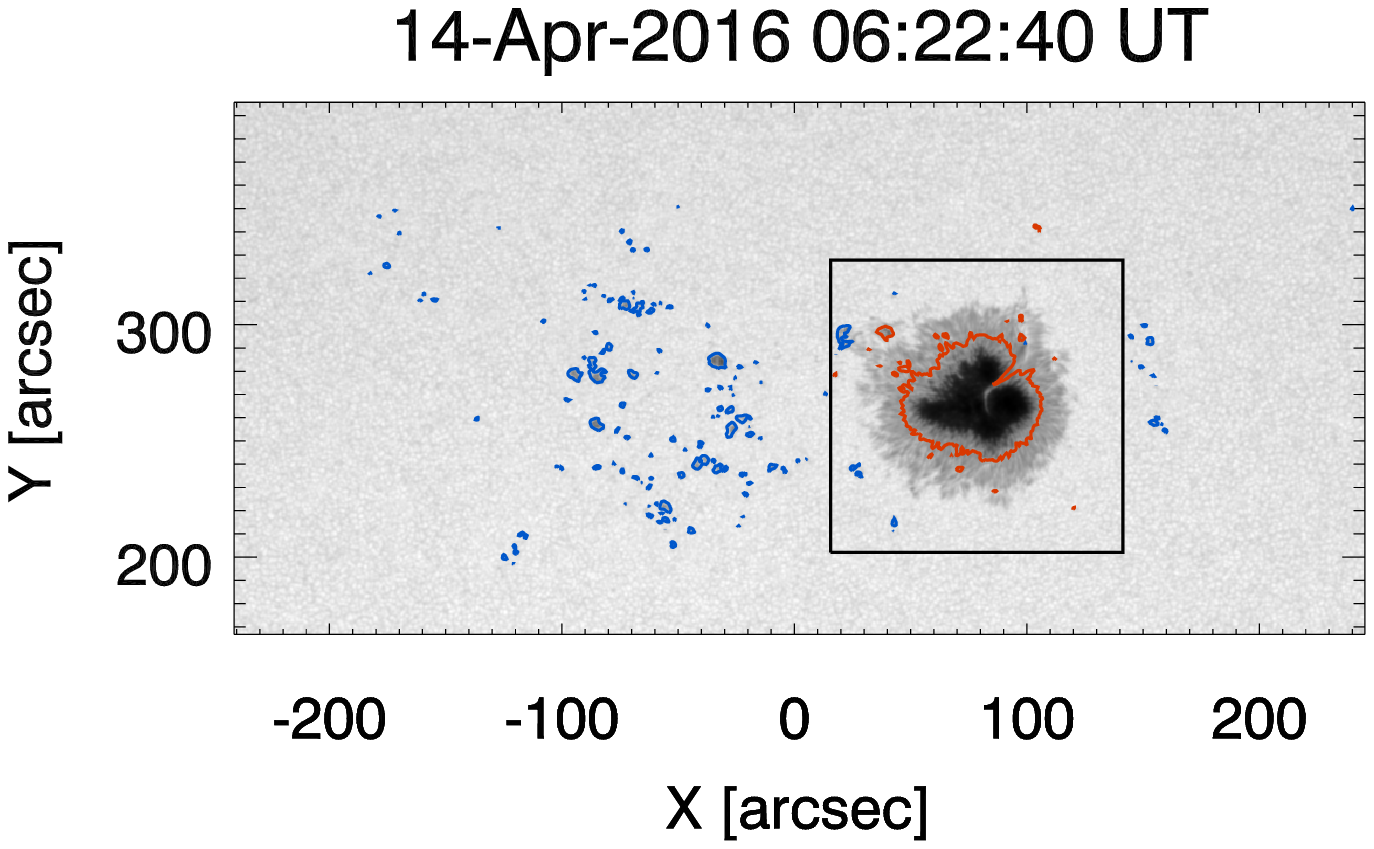}
	\caption{AR NOAA 12529 as seen in the continuum filtergram, with overlain contours of the simultaneous map of the vertical component of the magnetic field (red/blue = -800/+800~G), taken by HMI in the \ion{Fe}{1} 6173~\AA{} line, at the time of the central meridian passage of the AR. The box frames the portion of the field of view shown in Figure~\ref{fig2}. \label{fig1}}
\end{figure}

\section{Data and analysis}

We have analyzed Space-weather Active Region Patches (SHARPs) data \citep{Hoe14} acquired by the Helioseismic and Magnetic Imager \citep[HMI;][]{Sch12} onboard the \textit{Solar Dynamic Observatory} \citep[SDO;][]{Pes12} satellite from 2016 April 8 to April 19 to study the main sunspot of the AR NOAA 12529. We have used continuum filtergrams, Dopplergrams, and vector magnetograms acquired along the \ion{Fe}{1} line at 6173~\AA{}, with a pixel size of 0\farcs51 and a time cadence of 12~min.

The vector field has been computed using the Very Fast Inversion of the Stokes Vector code \citep{Bor11}, which performs a Milne-Eddington inversion of the observed Stokes profiles, optimized for the HMI pipeline; the remaining $180^{\circ}$-azimuth ambiguity is resolved with the minimum energy code \citep{Met94}. More details about the SHARP pipeline are reported in \citet{Bob14}. The SHARP data have been corrected for the rotation angle of $180^{\circ}$ of HMI images. Finally, the vector magnetic field components have been transformed into the local solar frame, according to \citet{Gar90}.

HMI Dopplergrams have been corrected for the effect of solar rotation, by subtracting the mean velocity averaged over 10 days, deduced from the HMI data series relevant to Carrington rotation 2176. Doppler velocity has been calibrated assuming umbra (i.e., pixels with normalized continuum intensity $I_c < 0.4$) on average at rest \citep[e.g.][]{Rim94}.

We have also used images acquired in the extreme ultraviolet (EUV) by the Atmospheric Imaging Assembly \citep[AIA;][]{Lem12} at 171 \AA{} and 304 \AA{}, with a pixel size of about 0\farcs6 and a time cadence of 12~s. 

Furthermore, we have analyzed full disc images of the chromosphere acquired in the H$\alpha$ line at 6562.8~\AA{} by INAF - Catania Astrophysical Observatory. These images have been coaligned together with HMI and AIA data.

\section{Results}

During the entire passage across the solar disc, AR NOAA 12529 was composed by a large preceding sunspot of negative polarity and some following pores of positive polarity, as shown in Figure~\ref{fig1}. The main sunspot was already formed when it appeared at the East limb on 2016 April 8, at heliographic latitude $\sim 10$~N. It occupied an area of about $2000 \,\mathrm{Mm}^2$.

\begin{figure*}
 \centering
	\includegraphics[trim=10 155 155 290, clip, scale=0.45]{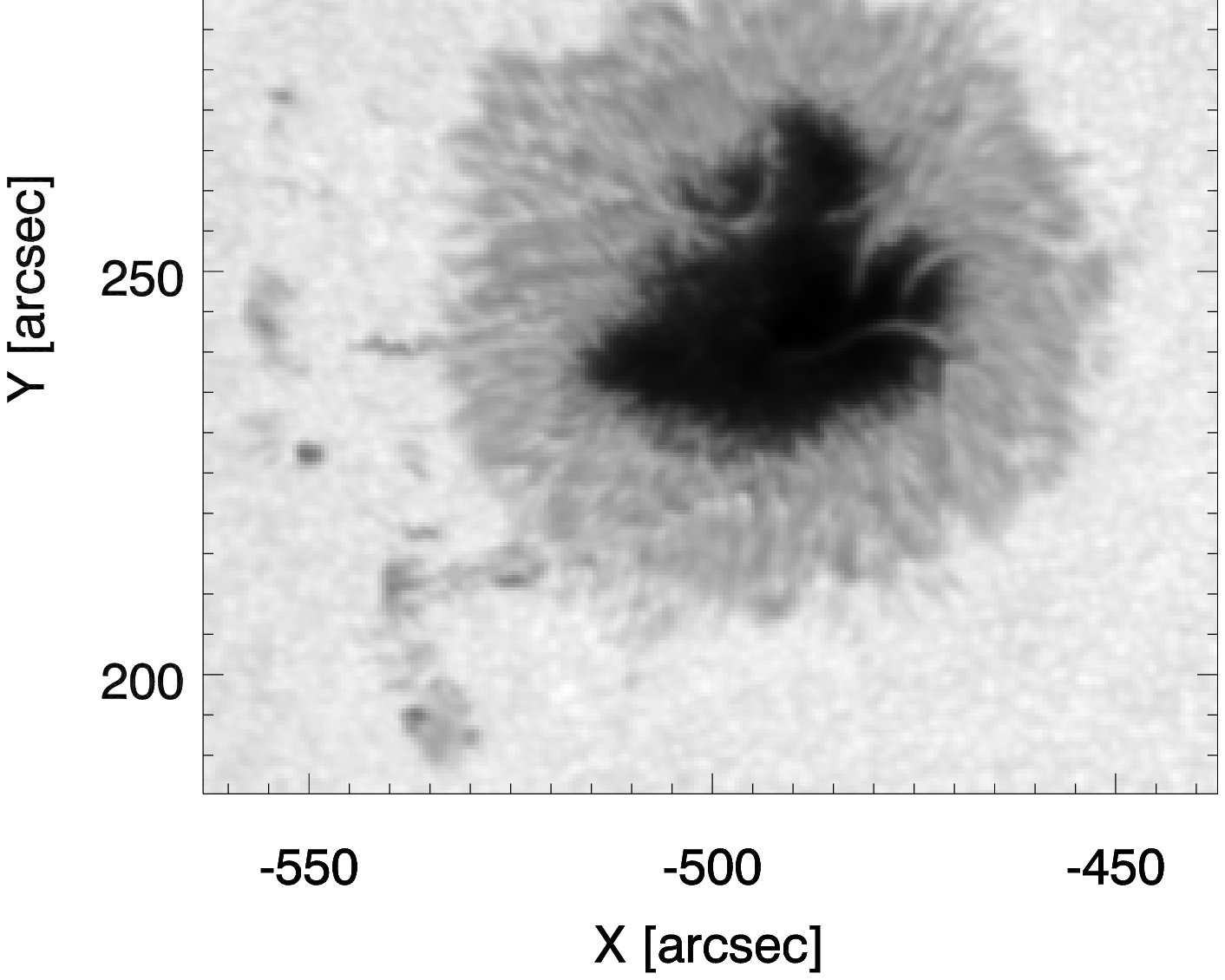}%
	\includegraphics[trim=40 155  40 290, clip, scale=0.45]{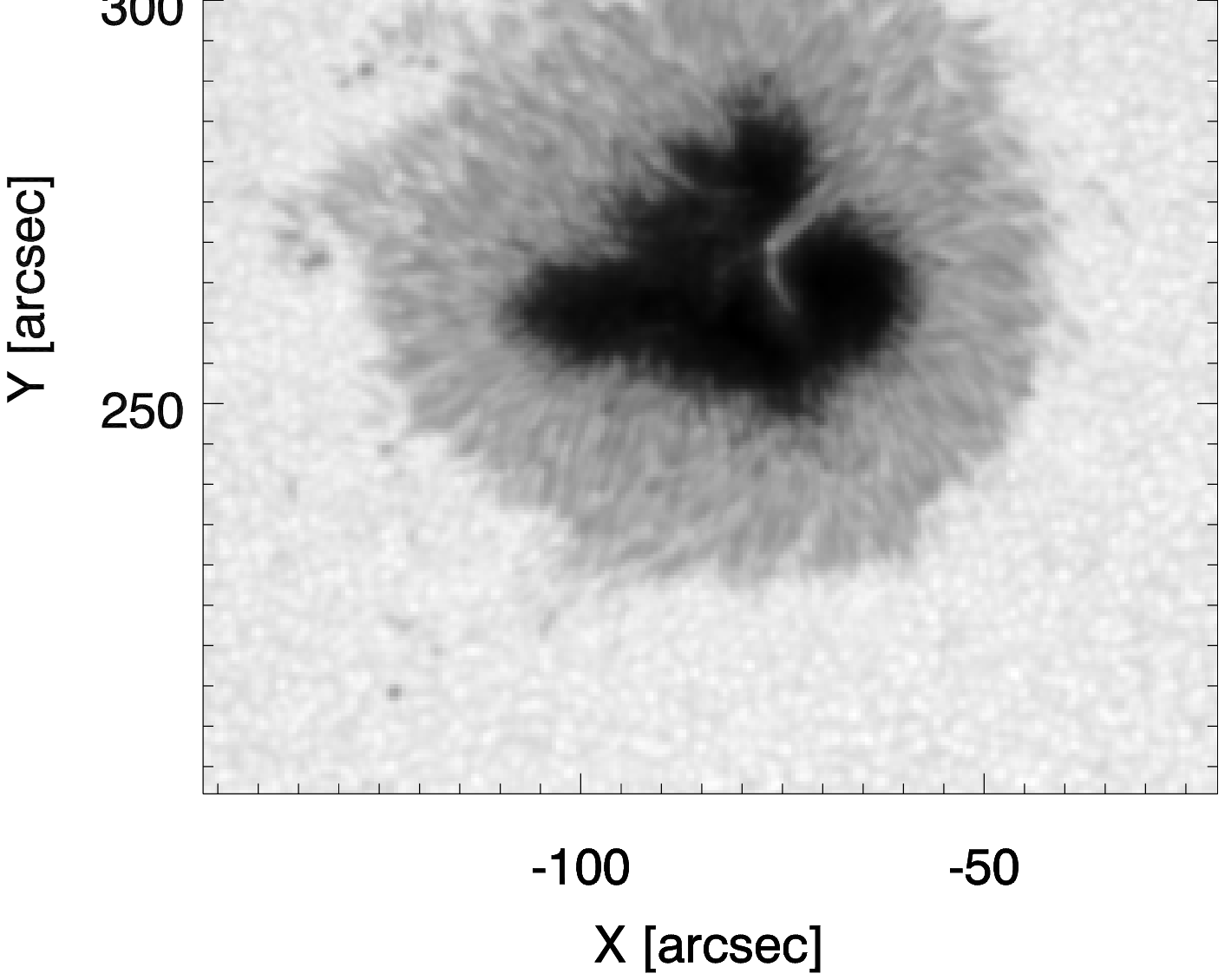}
	\includegraphics[trim=10 155 155 330, clip, scale=0.45]{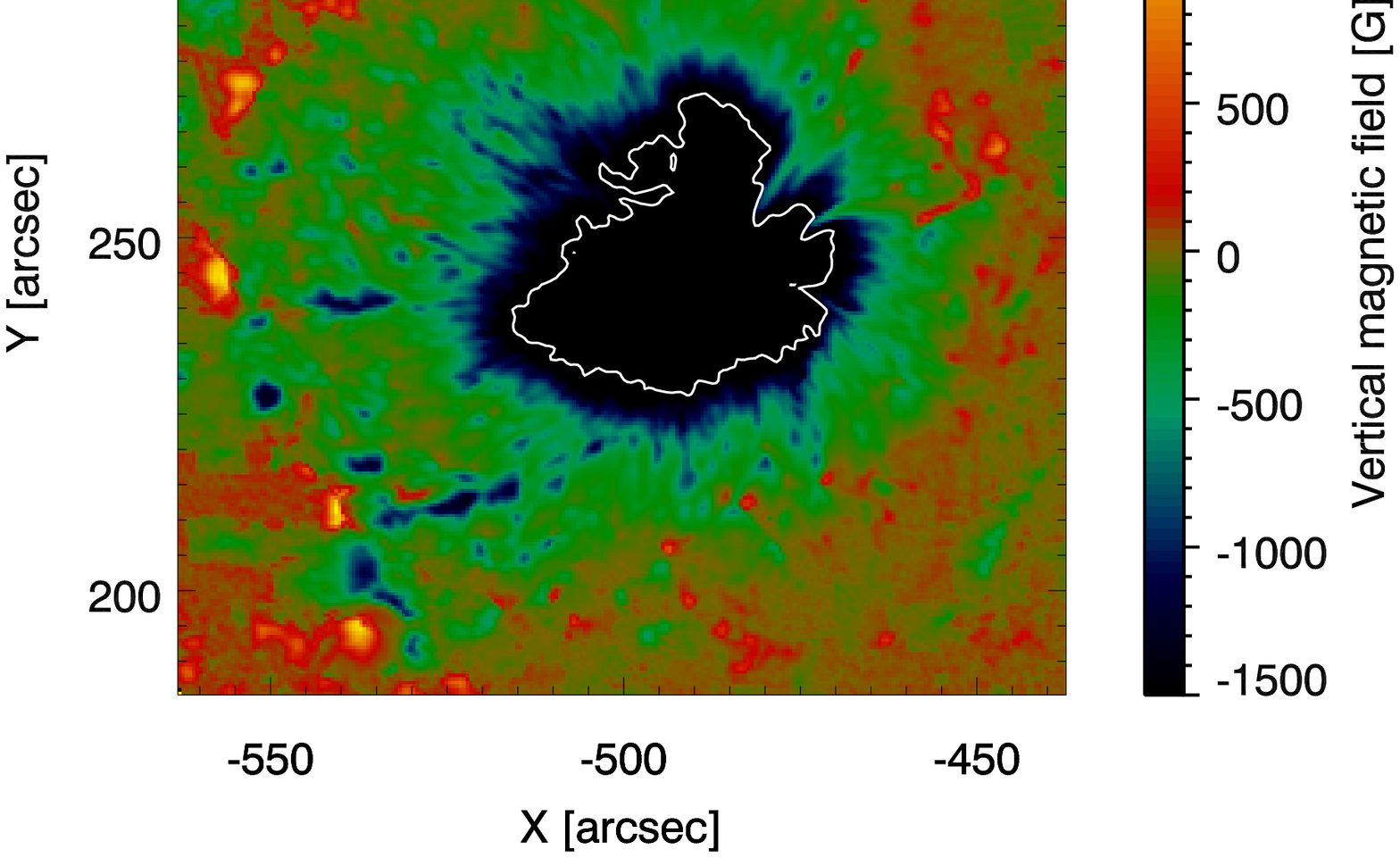}%
	\includegraphics[trim=40 155  40 330, clip, scale=0.45]{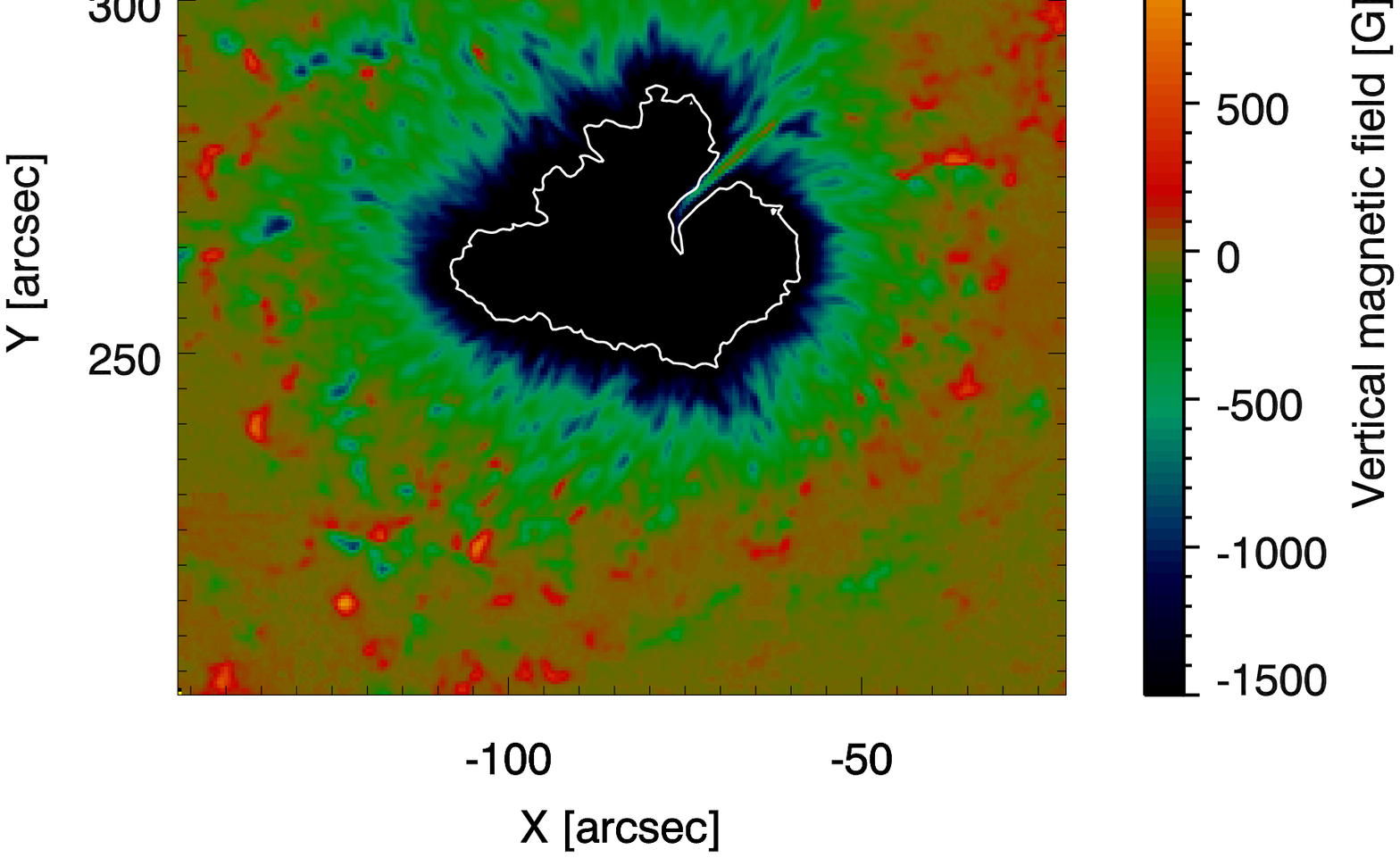}
    \includegraphics[trim=10 155 155 330, clip, scale=0.45]{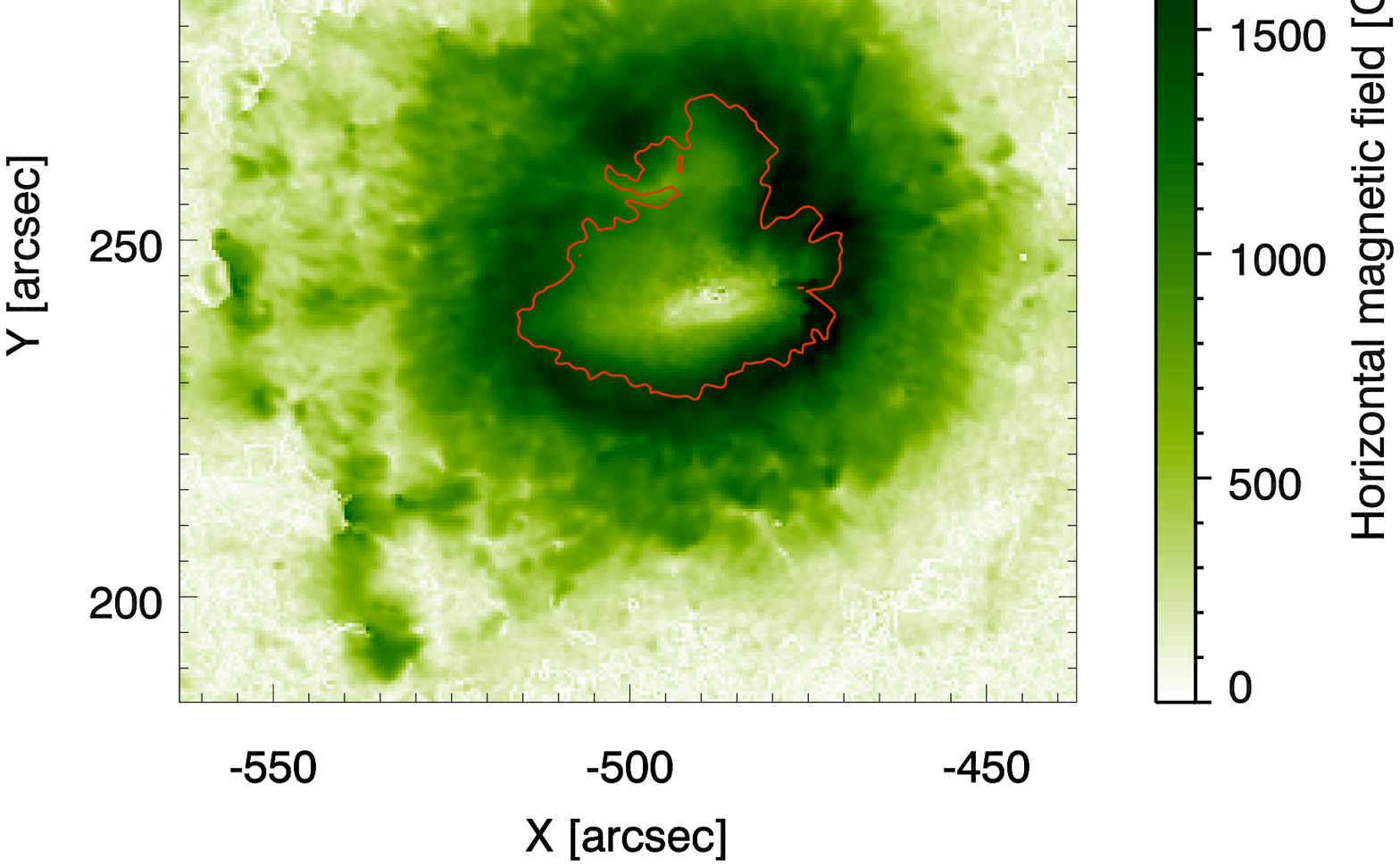}%
    \includegraphics[trim=40 155  40 330, clip, scale=0.45]{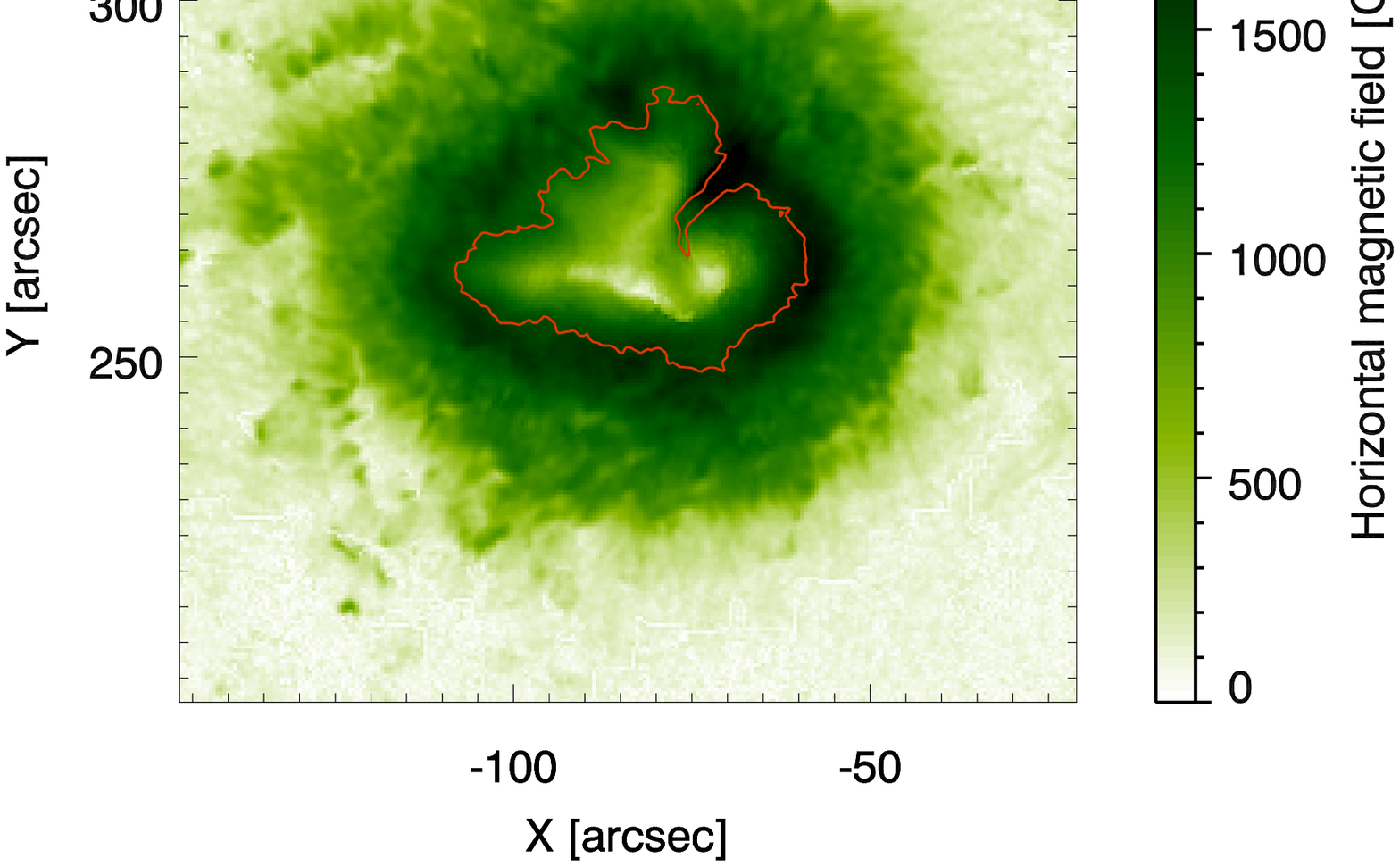}
    \includegraphics[trim=10  95 155 330, clip, scale=0.45]{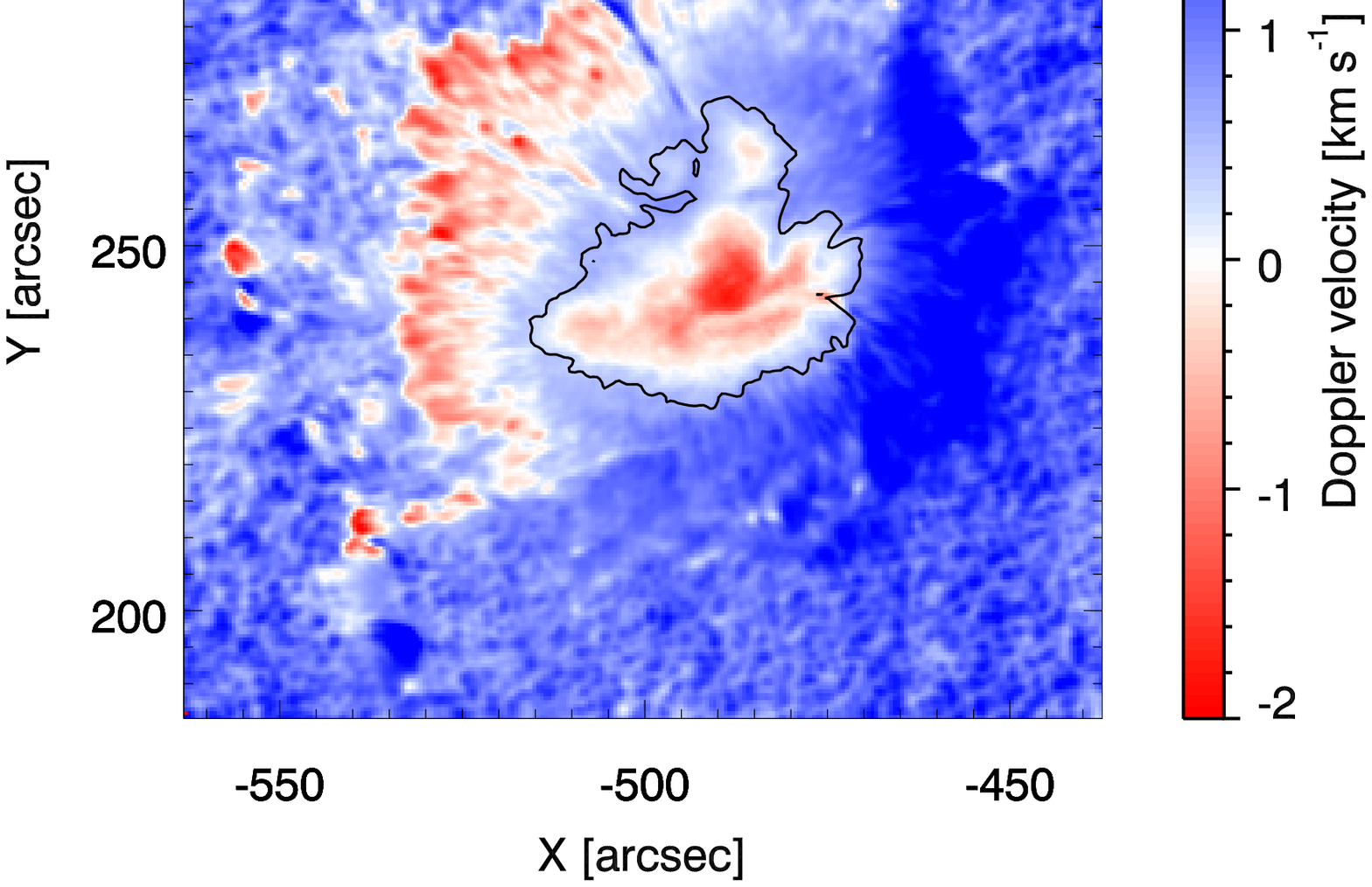}%
    \includegraphics[trim=40  95  40 330, clip, scale=0.45]{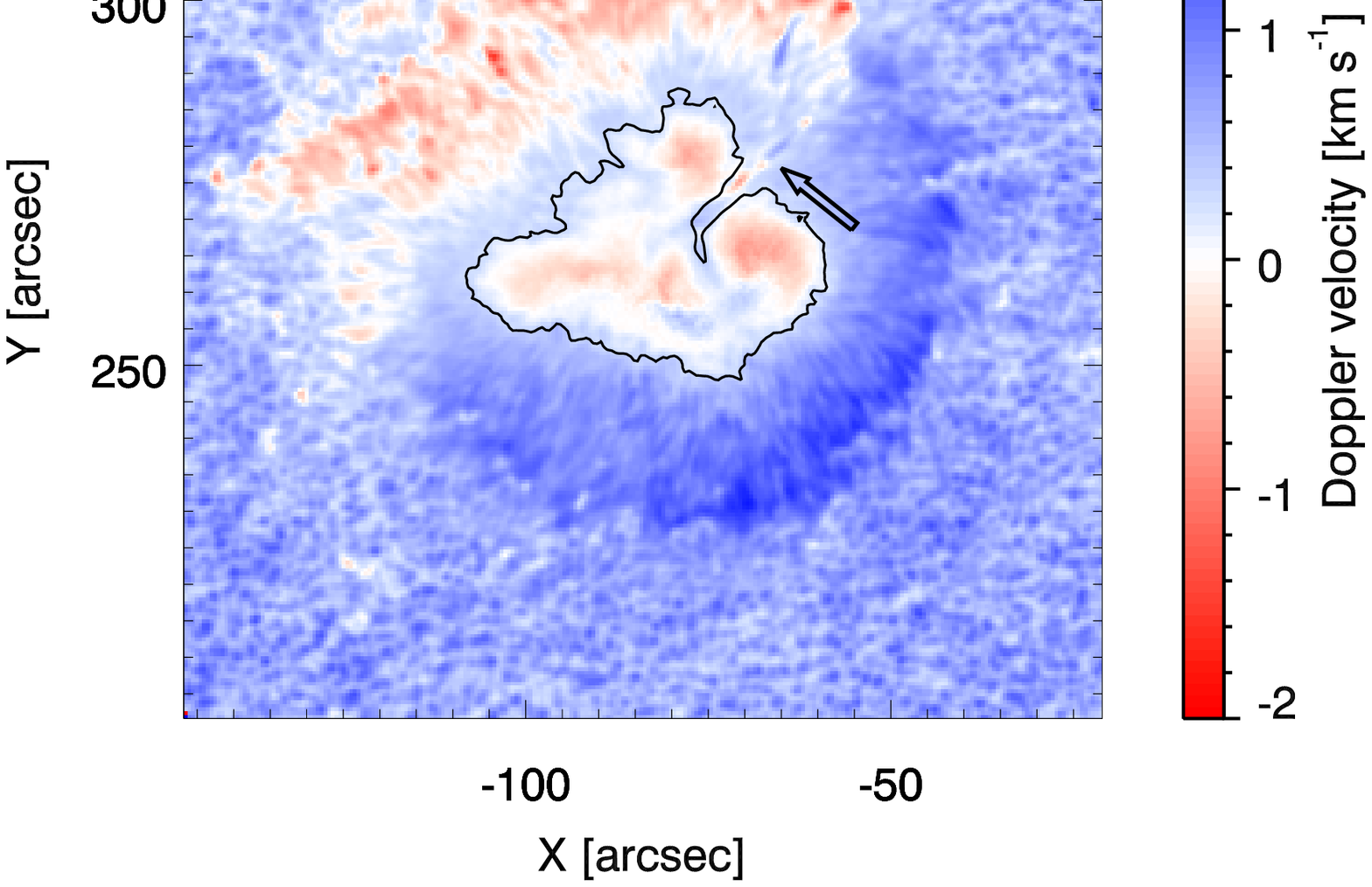}     
  \caption{Evolution of the main sunspot of AR NOAA 12529 between 2016 April 11 (left) and April 13 (right) as seen in the continuum filtergrams (first row), in the simultaneous maps of the vertical (second row) and horizontal (third row) magnetic field components, and in the simultaneous Dopplergrams (fourth row), taken by HMI in the \ion{Fe}{1} 6173~\AA{} line. Contours represent the umbra-penumbra boundary at $I_c = 0.4$. \label{fig2}}
\end{figure*}

On April 11, when the sunspot was located at $\mu = 0.8$, several bright elongated structures similar to small LBs appeared to the north-west of the spot, intruding into the umbra (see the top-left panel of Figure~\ref{fig2}). They have a curved shape, not showing a granular pattern at the spatial resolution of the HMI continuum images. In the map of the vertical component of the magnetic field taken at the same time (first column, second row of Figure~\ref{fig2}), we note that these structures are characterized by a lower field intensity, in comparison with the other parts of the sunspot umbra. The horizontal component of the magnetic field in the hook-shaped structure observed to the north-east of the sunspot, located at $[-500\arcsec,-490\arcsec] \times [250\arcsec, 260\arcsec]$, is weaker than the surrounding penumbral filaments, whereas in the intruding structures located to the west it is comparable to that of the penumbra (first column, third row of Figure~\ref{fig2}).

The longest of these structures, seen at $[-480\arcsec,-470\arcsec] \times [255\arcsec, 265\arcsec]$ on April 11, reached its maximum length of about 30\arcsec{} on April 13 (top-right panel of Figure~\ref{fig2}), when it was located at $[-80\arcsec,-65\arcsec] \times [65\arcsec, 80\arcsec]$, the sunspot being located at $\mu = 0.96$. At that time, a portion along the axis of the structure has a magnetic field of positive polarity, i.e., opposite to that of the magnetic field of the surroundings (second column, second row of Figure~\ref{fig2}). Moreover, the horizontal field in the structure is $\approx 2000 \,\mathrm{G}$, about 50\% stronger than in the surrounding penumbra (second column, third row of Figure~\ref{fig2}). Many of these bright structures, including the longest one, remained visible till the AR reached the West limb, when the sunspot appeared already fragmenting, in a decay phase. 

Interestingly, we find a noticeable behaviour of the plasma motions along the line of sight in these bright structures. On April 11, although the AR was near the East limb and the uncertainties due to the projection effects are not negligible, the velocity observed along the bright structures in the umbra (bottom-left panel of Figure~\ref{fig2}) is slightly larger than in the surroundings. In particular, the hook-shaped bright structure located to the north-eastern side of the umbra shows a velocity direction inverse to the Evershed flow in its northern leg. The bright structure to the western side of the umbra clearly also exhibits a plasma velocity higher than the surrounding Evershed flow, in this case along the normal Evershed flow direction. However, on April 13, when the longest of these structures reached its maximum length and the AR was approximating to the central meridian, we note plasma motions with both directions along it (see the arrow in the bottom-right panel of Figure~\ref{fig2}): upflows and downflows alternate from the outer to the inner part of the bright structure with velocities of the order of $1\,\mathrm{km \,s}^{-1}$. Each portion of the bright structure with the same direction of the plasma motion is about 5\arcsec{} long. Note that these alternate, coherent plasma flows may suggest a helical motion along the structure.

This behaviour is mostly visible when the sunspot passed near the central meridian, i.e., on April 13 and 14, but we cannot exclude that these plasma motions may persist during the whole life of the bright structure, i.e., for about 5 days, from April 11 to April 16, although they are not visible due to the foreshortening effects. 

\begin{figure}
 \centering
	\includegraphics[trim=10 125 120 480, clip, scale=0.6]{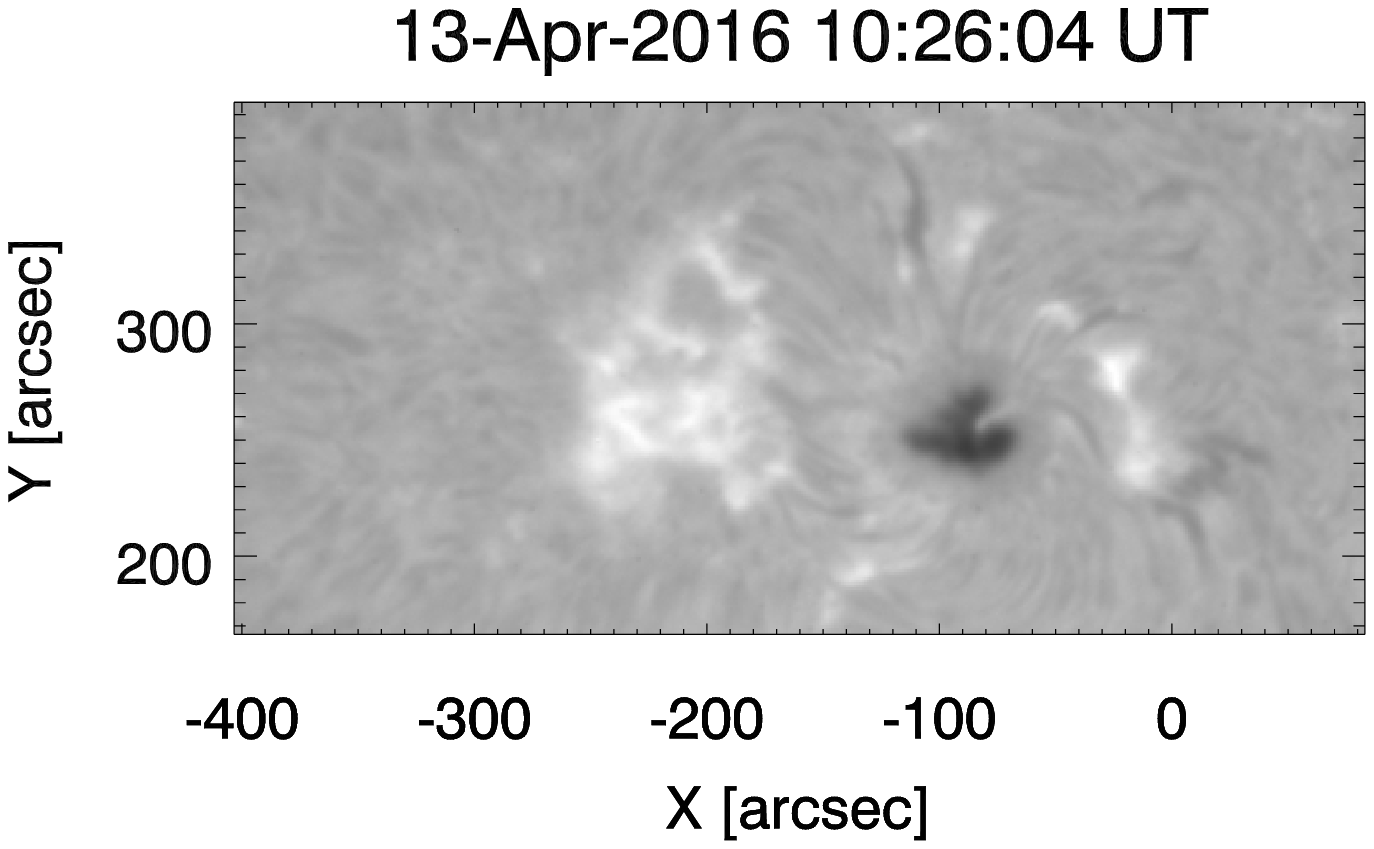}\\
	\includegraphics[trim=10 125 120 490, clip, scale=0.6]{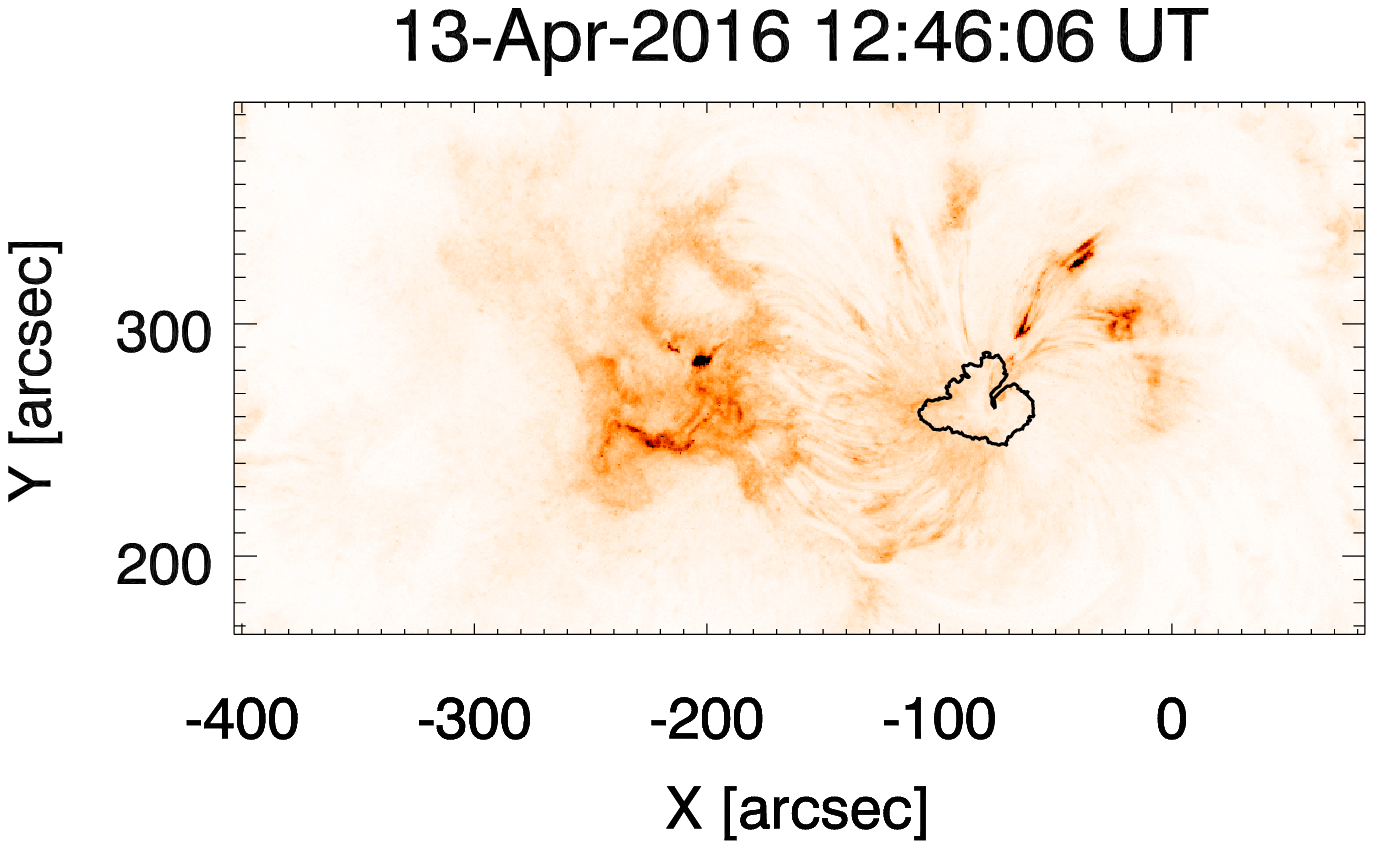}\\
    \includegraphics[trim=10  95 120 490, clip, scale=0.6]{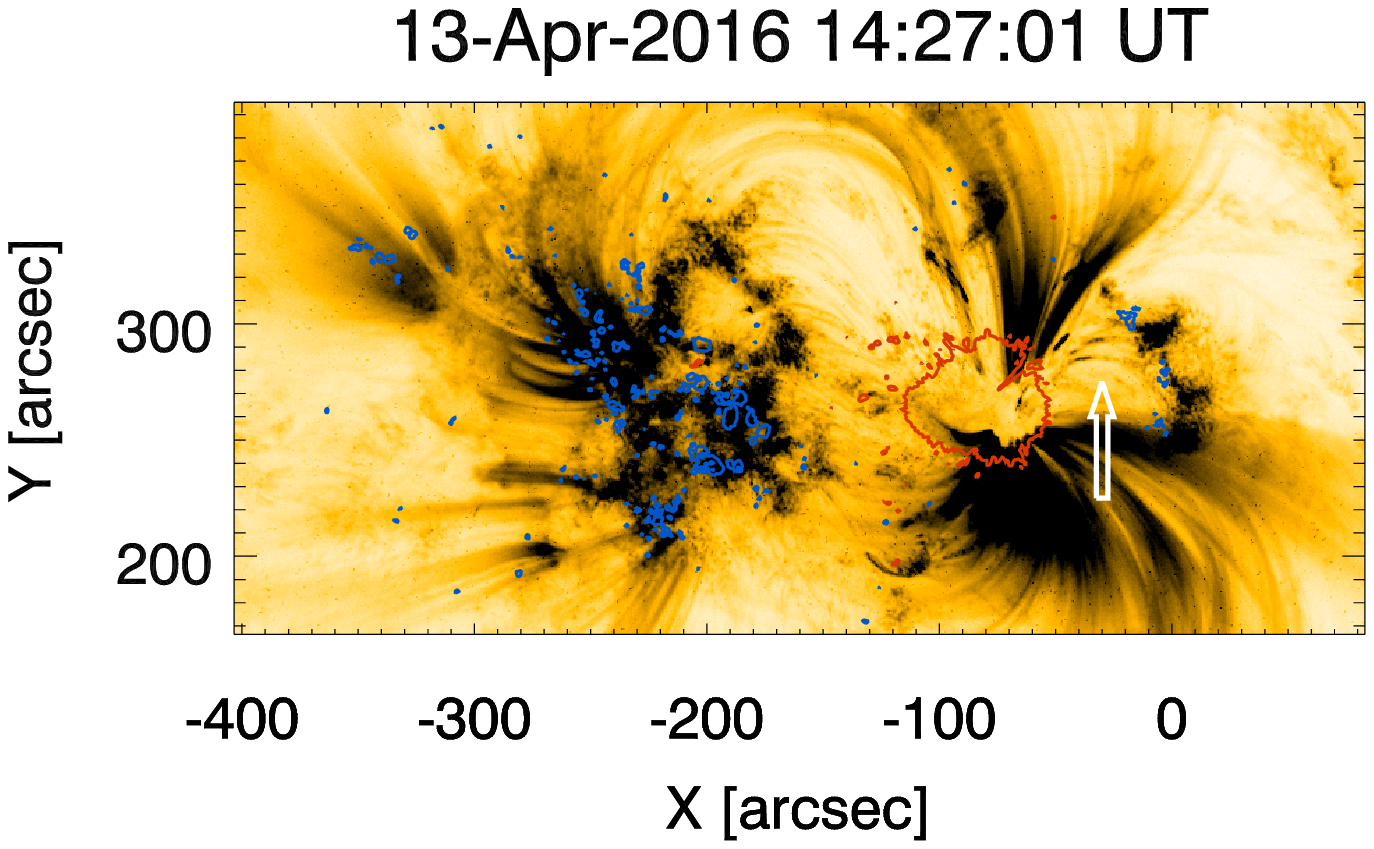}
  \caption{AR NOAA 12529 as seen in the H$\alpha$ line by INAF - Catania Astrophysical Observatory (top panel), at 304 \AA{} (middle panel) and at 171 \AA{} (bottom panel) by AIA. Note that AIA images are shown in a reversed color table. The contour in the middle panel corresponds to the umbra-penumbra boundary at $I_c = 0.4$ in the simultaneous continuum image. Red/blue contours in the bottom panel correspond to -400/+400~G in the simultaneous map of the vertical component of the magnetic field. The arrow in the bottom panel indicates the EUV filament channels corresponding to the filaments observed in the H$\alpha$ line and at 304~\AA. \label{fig3}}
\end{figure}

In the image of the chromosphere acquired on April 13 at 10:26 UT in the H$\alpha$ line at INAF -- Catania Astrophysical Observatory, we note the presence of chromospheric filaments located to the north-western side of the sunspot (top panel of Figure~\ref{fig3}). These filaments are also visible  in the AIA image at 304~\AA{} (middle panel of Figure~\ref{fig3}), connecting the sunspot to opposite polarity flux concentrations observed to the west of the preceding spot. It is worth noting that these filaments have the same curvature as the longest photospheric bright structure observed in the umbra of the sunspot. In particular, the portion of the filament corresponding to the bright structure inside the umbra at photospheric level also appears bright at 304~\AA{}. Moreover, at 171~\AA{} the filaments correspond to EUV filament channels (see the bottom panel of Figure~\ref{fig3}), surrounded by some bright loops with the same curvature, indicated with an arrow in the same figure.

\section{Discussion and conclusions}

In April 2016 a big sunspot passed over the visible hemisphere of the Sun showing bright elongated features apparently similar to unsegmented LBs. In this Letter we analyzed these structures, in particular the longest one, whose characteristics suggest a completely different origin with respect to LBs. 

We found that no granular pattern was visible in this intruding filamentary feature at the HMI resolution. Moreover, the magnetic field strength in the structure is not weaker than in the surrounding umbral region. At the same time, we showed the presence of a strong horizontal field component of the magnetic field in the structure and a vertical component opposite to that of the surrounding penumbral filaments in a portion of it. Peculiar line-of-sight plasma motions along this structure are also observed at the photospheric level. In addition, images of the chromosphere at 6562.8~\AA{} and at 304~\AA{} reveal the existence of some small filaments in the nearby, which are characterized by the same curvature of the photospheric structure, corresponding to EUV channels observed at 171~\AA{}.

As regards the magnetic configuration, one has to be cautious with the interpretation of the results deduced from the Milne-Eddington solution implemented in the VFISV code. In fact, it may not account for the complexity of the Stokes profiles present in regions where peculiar processes are at work. Actually, this affects the values of $\chi^2$ of the fits, which are $2-3$ times higher in the bright structure than in the spot. Moreover, the smearing due to the point-spread function of the telescope may dilute into the bright structure part of the magnetic signal belonging to the surrounding umbra and penumbral filaments. However, the average estimated errors for the vertical and horizontal component of the field in the structure are about $\pm 35$~G and $\pm 70$~G, respectively. Thus, the strong horizontal field of about 2000~G appears to be a robust result.

The presence of such a significant horizontal component of the magnetic field allows us to exclude the field-free configuration typically observed in LBs. Conversely, it supports the possibility that the magnetic field of the flux rope forming the filaments around the sunspot may have a counterpart in the middle of the sunspot umbra, where it appears to end in the cospatial chromospheric images. This would also provide an explanation for the detection of opposite polarity in the structure with respect to the umbra, as well as an alternative interpretation for the opposite polarity of an apparently filamentary LB with respect its hosting umbra observed by \citet{Bha07}. Note that this phenomenon is different from the magnetic reversals found by \citet{Lag14} and \citet{Fel16} in LBs, due to the bending of the field lines along the atmospheric stratification of the structure.

Moreover, this scenario is confirmed by the portions of the bright feature characterized by both upward and downward plasma motions along it, which are compatible with plasma flows along the helical field lines of a flux rope. Actually, the length of each portion of the filamentary structure with coherent plasma motion is of about 5\arcsec{}, which could correspond to the pitch of a helically twisted flux tube. 

Therefore, we guess that this bright filamentary structure is not a filamentary LB, but it is due to the accumulation of plasma coming from higher solar atmospheric levels into the photosphere. The presence of stably inclined fields belonging to the flux rope, which are touching part of the sunspot umbra, would set in radiatively driven penumbral magneto-convective mode \citep{Jur14, Jur17}. Indeed, high-resolution observations have shown that plasma fallen from chromospheric  layers can lead to the formation of penumbra in sunspots \citep{Shi12, Rom13, Rom14}. Furthermore, penumbral-like structures, i.e., so called orphan penumbrae, have been reported as a photospheric counterpart of a chromospheric filament \citep{Kuc12a, Kuc12b, Bue16}.

Observations of similar bright filaments inside the umbra, called UFs, have been recently reported by \citet{Kle13}. They found that UFs were characterized by a horizontal flow opposite to the Evershed flow and they also found from coronal images the presence of bright coronal loops which seemed to end in UFs. To interpret such phenomena, \citet{Kle13} conjectured two different scenarios. In the first, the bright filament ending in the umbra is formed by a sheet spanning many atmospheric layers and producing a siphon inflow by the pressure difference between the umbra and the network. In this case, the emission observed in the filament could be due to the energy dissipation at the boundary layers between the sheet and the sunspot magnetic field. In the second scenario, they interpreted the emission as an effect of a thick magnetic flux tube with high enough density such that the observed region is formed higher in the atmosphere. 

However, in both cases \citet{Kle13} did not take into account the presence of a flux rope whose signatures are clearly visible in our observation, albeit the second scenario they proposed shares some similarities with ours. In fact, we do not detect counter Evershed flow, with the exception of one leg of the hook-shaped bright structure observed on April 11 to the north-eastern side of the umbra. We find, instead, a peculiar pattern of the plasma flows which seems to describe a helical motion along the flux rope of a filament ending in the sunspot umbra. Note that the target of \citet{Kle13} suffered from severe projection effect, due to the position of AR NOAA 11302 on the solar disc when UFs were observed ($\mu \approx 0.6$).

It is also worth mentioning that AR NOAA 12529 was not a flare-rich AR: the strongest event was a unique M-class (M6.7) flare on April 18, while \citet{Kle13} suggested that the UFs could be related to high flare-productivity, as they would influence the structure of the overlying coronal loop leading to magnetic rearrangement. 

To support our scenario, further analyses are necessary. In this respect, in a next paper we shall analyze the high-resolution observations performed by the spectropolarimeter aboard the \textit{Hinode} satellite in the photosphere and by the \textit{IRIS} spacecraft during the central meridian passage of AR NOAA 12529. A preliminary analysis of the \textit{Hinode} data shows that a strong linear polarization signal is detected in the structure within the umbra, which also exhibits a portion with circular polarization of sign opposite to that of the surrounding umbra, much larger than that observed with HMI. Note that these polarization signals cannot be due to the contamination induced by the point spread function of the telescope with the signals from the surrounding magnetic elements. This strengthens the interpretation that we are proposing here.

The question about the conditions to observe such a phenomenon remains still open. Why do we not often observe this kind of bright structures similar to LBs but without any granular pattern? Can these structures be caused by the particular proximity of the footpoints of chromospheric filaments to a sunspot? Or can they be due to a peculiar configuration of the flux rope forming these filaments? An answer will likely be obtained by benefitting from the high spatial resolution and continuous temporal coverage provided by the next generation of solar observatories, such as the Solar Orbiter space mission \citep{Muller:13} and the large-aperture ground-based telescopes DKIST \citep{Keil:10} and EST \citep{Collados:10}.



\acknowledgments

The authors wish to thank an anonymous referee for his/her insightful comments. This work was supported by the Istituto Nazionale di Astrofisica (PRIN-INAF 2014), by the Italian MIUR-PRIN grant 2012P2HRCR on The active Sun and its effects on space and Earth climate, by Space Weather Italian COmmunity (SWICO) Research Program, and by the Universit\`a degli Studi di Catania. The research leading to these results has received funding from the European Union's Horizon 2020 research and innovation programme under grant agreement no.~739500 (PRE-EST project). The SDO data used in this paper are courtesy of NASA/SDO science team. 

\facility{SDO (HMI, AIA)}, \facility{OACT:0.15m}

\end{document}